\documentclass[a4paper]{amsart}
\usepackage{fullpage}

	\usepackage{amsfonts,amssymb,amsmath}
	\usepackage{amsmath,amsthm}
	\usepackage[mathscr]{euscript}

			\usepackage[usenames,dvipsnames]{color}
			\usepackage{soul}
			\sethlcolor{cyan} 
			\newcommand{\myIdentity}[1][]{
				\medskip
				\noindent
				\ul{Identity #1.}\ 
			}
			\newcommand{\myTheorem}[1][]{
				\medskip
				\noindent
				\underline{Theorem.}\ 
			}

	\usepackage[dvips]{pstricks}
	\psset{xunit=.5pt,yunit=.5pt,runit=.5pt}

\begin{document}

\title[$F$-matrices and XXZ spin-$\frac{1}{2}$ chain]{Factorizing $F$-matrices and the XXZ spin-$\frac{1}{2}$ chain \\ --- A diagrammatic perspective ---}

\author{S G M\textsuperscript{c}Ateer and M Wheeler}

\address{Department of Mathematics and Statistics,
                 The University of Melbourne,
                 Parkville, Victoria, Australia.}
\email{s.mcateer@ms.unimelb.edu.au, m.wheeler@ms.unimelb.edu.au}


\begin{abstract}
Using notation inherited from the six-vertex model, we construct diagrams that represent the action of the factorizing $F$-matrices associated to the finite length XXZ spin-$\frac{1}{2}$ chain. We prove that these $F$-matrices factorize the tensor $R^{\sigma}_{1\ldots n}$ corresponding with elements of the permutation group. We consider in particular the diagram for the tensor  $R^{\sigma_c}_{1\ldots n}$, which cyclically permutes the spin chain. This leads us to a diagrammatic construction of the local spin operators $S_i^{\pm}$ and  $S_i^{z}$ in terms of the monodromy matrix operators.
\end{abstract}

\maketitle

\setcounter{section}{0}

\section{Introduction}\label{intro}

The idea of {\it twisting} in quantum groups \cite{dri1,dri2,dri3} was applied in \cite{ms} in the context of the algebraic Bethe Ansatz, where in this case the twist is represented by an $F$-matrix $F_{1\ldots n}$ which acts in the tensor product of quantum spaces $V_1\otimes \cdots \otimes V_n$ and satisfies the factorizing equation
\begin{align}
	F_{\sigma(1)\ldots \sigma(n)}
	R^{\sigma}_{1\ldots n}
	=
	F_{1\ldots n}
	\label{fact-big}
\end{align}
where $\sigma \in S_n$ is an arbitrary permutation, and $R^{\sigma}_{1\ldots n}$ represents a tensor product of $R$-matrices associated to $\sigma$. Since $F_{\sigma(1)\ldots \sigma(n)}$ is invertible we have
\begin{align}
	R^{\sigma}_{1\ldots n}
	=
	(F_{\sigma(1)\ldots \sigma(n)})^{-1}
	F_{1\ldots n},
\end{align}
hence $F_{1\ldots n}$ is referred to as a \emph{factorizing} $F$-matrix.  One of the main results in \cite{ms} was an explicit formula for $F_{1\ldots n}$ when specializing to representations of the quantum affine algebra $U_q(\widehat{sl_2})$. This expression for the $F$-matrix is entirely in terms of products of the corresponding trigonometric $R$-matrix. 

The aim of this paper is to reproduce several of the results from \cite{ms}, using diagrammatic tensor notation along the lines of \cite{pen}. In particular we give a diagrammatic representation of the $F$-matrix $F_{1\ldots n}$, using conventions from the six-vertex model to draw it as a lattice of vertices. This representation allows us to establish identities, such as (\ref{fact-big}), virtually by inspection.

A key aspect of our approach is a notational trick concerning the crossing of lattice lines. To illustrate this device, we consider the simplest factorizing equation which is obtained from (\ref{fact-big}) by setting $n=2$ and $\sigma\{1,2\} = \{2,1\}$,
\begin{align}
	F_{21} R_{12} = F_{12}
	\label{fact-small}
\end{align}
where $R_{12} \in {\rm End}(V_1 \otimes V_2)$ represents the trigonometric $R$-matrix associated to $U_q(\widehat{sl_2})$. As was noticed in \cite{ms}, a solution of this equation is given by 
\begin{align}
	F_{12}
	=  
	e^{(++)}_1 I_{12} + e^{(--)}_1 R_{12}
	=
	e^{(--)}_2 I_{12} + e^{(++)}_2 R_{12}
	\label{f1}
\end{align}
where $I_{12} \in {\rm End}(V_1 \otimes V_2)$ is the identity tensor, while $e^{(i j)}_k \in {\rm End}(V_k)$ is the $2\times2$ matrix with entry $(i,j)$ equal to 1 and all remaining entries equal to 0.\footnote{To eliminate confusion between component labels and the labels assigned to vector spaces, we label the components of a $2\times 2$ matrix by $\{+1,-1\} \equiv \{+,-\}$ rather than $\{1,2\}$.} A suitable diagrammatic representation of the $F$-matrix (\ref{f1}) is obtained as follows. Using the standard vertex notation for the components of the $R$-matrix,
\begin{align}
(R_{12})^{i_1 j_1}_{i_2 j_2} &\ \input{diagrams/define-r-tensor}
	\label{r1}
\end{align}
where $i_1,i_2,j_1,j_2$ take values in $\{+,-\}$, the line labelled by $i_k,j_k$ can be said to correspond with the vector space $V_k$. The crossing of the two lines indicates the action of $R_{12}$  in $V_1 \otimes V_2$. On the other hand, the components of the identity tensor $I_{12} \in {\rm End}(V_1 \otimes V_2)$ can be represented as
\begin{align}
	(I_{12})^{i_1 j_1}_{i_2 j_2}&\ \input{diagrams/define-identity-tensor-2}. 
	\label{i3}
\end{align}
Now the lines are considered to be decoupled, indicating the trivial action of $I_{12}$ in $V_1 \otimes V_2$. Based on the notations (\ref{r1}) and (\ref{i3}), we represent the components of the $F$-matrix (\ref{f1}) by
\begin{align}
	(F_{12})^{i_1 j_1}_{i_2 j_2} &\ \input{diagrams/define-f-tensor-1}
	\label{f2}
\end{align}
The dot $\bullet$ which has been placed at the center of the vertices (\ref{f2}) denotes that the crossing of the lines 1 and 2 is sensitive to the value of the index next to the open-faced arrows $\triangledown,\vartriangle$. In the vertex on the left, when $i_1 = +$ the lines \emph{do not} cross and when $i_1=-$ they \emph{do} cross. In the vertex on the right, when $i_2=+$ the lines \emph{do} cross and when $i_2 = -$ they \emph{do not} cross.\footnote{i.e. when we identify $+$ with $\uparrow$ and $-$ with $\downarrow$, \emph{agreement} with the open-faced arrow corresponds to \emph{crossing}, \emph{disagreement} to \emph{non-crossing}.} It is easy to check that the components recovered from (\ref{f2}) match the components obtained from equation (\ref{f1}). This crossing/uncrossing device underpins the rest of our work, and will be expanded upon in \S \ref{3.1}. 

The role of Drinfel'd twists in the algebraic Bethe Ansatz was further considered in \cite{kmt}. There it was noticed that the calculation of certain objects in the XXZ spin-$\frac{1}{2}$ chain, such as the domain wall partition function and scalar product, is significantly simplified when the objects in question have been transformed appropriately using $F$-matrices. In the same paper the authors also found formulae for the local spin operators $S_i^{\pm}$ and $S_i^{z}$ entirely in terms of global, monodromy matrix operators. This was achieved using the total symmetry of the monodromy matrix in the $F$-basis, and equation (\ref{fact-big}) in the case of a cyclic permutation $\sigma_c$. The second aim of this paper is to reproduce these formulae from a diagrammatic standpoint. It turns out that knowledge of $F$-matrices is {\it not} necessary to perform this calculation, a fact which was also observed in \cite{gk}, \cite{mt}.

In \S 2 we recall the six-vertex notation for the $U_q(\widehat{sl_2})$ $R$-matrix. A number of standard identities such as the unitarity, Yang-Baxter and intertwining equations are presented in diagrammatic form. We also give a diagrammatic interpretation of $R_{1\ldots n}^{\sigma}$ as a bipartite graph. In \S 3 we provide a diagrammatic exposition of the paper \cite{ms}. We use the notational device (\ref{f2}) to construct {\it partial} $F$-matrices, and give proofs of some basic identities in \cite{ms}. Attaching the partial $F$-matrices in a systematic fashion, we obtain the full $F$-matrix and show that it satisfies (\ref{fact-big}). In \S 4 we list the formulae obtained in \cite{kmt} for the reconstruction of the local spin operators $S_i^{\pm}$ and $S_i^{z}$. Starting from the diagrammatic representation of the cyclic propagator $R^{\sigma_c}_{1\ldots n}$, we outline a purely diagrammatic proof of these results.

\section{Vertex notation for trigonometric $R$-matrix}

\subsection{Trigonometric $R$-matrix $R_{12}$}\label{2.1}

An essential object in the quantum inverse scattering/algebraic Bethe Ansatz scheme is the quantum $R$-matrix. The $R$-matrix associated to the XXZ spin-$\frac{1}{2}$ chain is given by  
\begin{align}
	R_{12}(\xi_1,\xi_2)
	=
	\left(
	\begin{array}{cccc}
	a(\xi_1,\xi_2) & 0 & 0 & 0
	\\
	0 
	& 
	b(\xi_1,\xi_2) 
	& 
	c(\xi_1,\xi_2)
	& 
	0
	\\
	0
	&
	c(\xi_1,\xi_2)
	&
	b(\xi_1,\xi_2) 
	&
	0
	\\
	0 & 0 & 0 & a(\xi_1,\xi_2)
	\end{array}
	\right)_{12}
	\label{r2}
\end{align}
where the entries of (\ref{r2}) are the trigonometric functions
\begin{align}
	a(\xi_1,\xi_2) = 1,
	\quad
	b(\xi_1,\xi_2) =
	\frac{\sinh(\xi_1-\xi_2)}{\sinh(\xi_1-\xi_2+\eta)},
	\quad
	c(\xi_1,\xi_2) =
	\frac{\sinh\eta}{\sinh(\xi_1-\xi_2+\eta)}
	\label{weights}
\end{align}
and $\xi_1,\xi_2$ are free variables, while $\eta$ is the crossing parameter of the model. The matrix (\ref{r2}) is a representation of the $U_q(\widehat{sl_2})$ $R$-matrix on the tensor product $V_1\otimes V_2$, where $V_1,V_2$ are copies of $\mathbb{C}^2$. In order to denote this fact the matrix (\ref{r2}) is given the subscript $12$, and its variables are accordingly $\xi_1,\xi_2$. Often, we will abbreviate $R_{12}(\xi_1,\xi_2)$ to $R_{12}$ when the variables of the $R$-matrix are clear from context.

The $R$-matrix is a rank 4 tensor whose components have been arranged in a two-dimensional array for convenience. Correspondingly, four indices are needed to label the components of (\ref{r2}):   
\begin{align}
	R_{12}(\xi_1,\xi_2)
	=
	\left(
	\begin{array}{cccc}
	R^{++}_{++} & R^{++}_{+-} & R^{+-}_{++} & R^{+-}_{+-}
	\\
	R^{++}_{-+} & R^{++}_{--} & R^{+-}_{-+} & R^{+-}_{--}
	\\
	R^{-+}_{++} & R^{-+}_{+-} & R^{--}_{++} & R^{--}_{+-}
	\\
	R^{-+}_{-+} & R^{-+}_{--} & R^{--}_{-+} & R^{--}_{--}
	\end{array}
	\right)_{12}\ .
\end{align}
For all $i_1,i_2,j_1,j_2 \in \{+,-\}$ we can then define
\begin{align}
	(R_{12})^{i_1 j_1}_{i_2 j_2} &\ \input{diagrams/define-r-tensor-2}\ .
	\label{r-vert}
\end{align}
The object on the right hand side of (\ref{r-vert}) is called an $R$-{\it vertex}.\footnote{We have drawn two vertices on the right hand side of (\ref{r-vert}), because we use both notations interchangeably. In situations where we employ the latter notation, the reader should remember that each number actually stands for an {\it index} subscripted by that number. Numbers at the base of a diagram stand for indices of type $i$, while numbers at the top of a diagram stand for indices of type $j$. Notice that each label also encodes the rapidity flowing through that particular line.  Also note that the orientation of each vertex is fixed by setting the direction of rapidity flow to be \emph{up the page}; each vertex will have two lines which exit at the top of the diagram and two which exit at the bottom.}  We call these $R$-vertices to distinguish them from the $P$-vertices and $I$-vertices that we define below.  The indices placed on the four bonds of the vertex are called {\it state variables.} The state variable $+(-)$ is interchangeable with an arrow pointing up (down) the page. In this notation, the non-zero entries of (\ref{r2}) get represented by vertices possessing {\it arrow conservation.} This means that the number of upward arrows entering the base of the vertex is equal to the number of upward arrows leaving the top of the vertex. Only six configurations of arrows obey this property, leading to the equations
\begin{align}
	\input{diagrams/a-vertices} &\ a(\xi_1,\xi_2) \\
	\input{diagrams/b-vertices} &\ b(\xi_1,\xi_2) \\
\input{diagrams/c-vertices} &\ c(\xi_1,\xi_2)
\end{align}
while all remaining vertices are equal to zero. This achieves the well known correspondence between the XXZ spin-$\frac{1}{2}$ chain and the six-vertex model of statistical mechanics.

Further, the notation (\ref{r-vert}) allows a succinct representation of tensor multiplication (contraction). For example, the components of the rank 6 tensor $R_{13}(\xi_1,\xi_3)R_{12}(\xi_1,\xi_2)$ are represented by the diagram
\begin{align}
	(R_{13})^{i_1 k_1}_{i_3 j_3}
	(R_{12})^{k_1 j_1}_{i_2 j_2}
	&\ \input{diagrams/contraction}
	\label{r-vert2}
\end{align}
where summation is implied over the repeated index $k_1$. The label $k_1$ in the diagram (\ref{r-vert2}) is superfluous if we adopt the convention that all internal bonds are summed over. In the sequel, we represent tensors of rank $2n$ by diagrams possessing $2n$ external bonds, while all internal bonds are implicitly summed. 

\subsection{Permutation matrix $P_{12}$}
\label{perm}

In the case where the variables $\xi_1,\xi_2$ are equal, the $R$-matrix (\ref{r2}) satisfies
\begin{align}
	R_{12}(\xi,\xi)
	=
	\left(
	\begin{array}{cccc}
	1 & 0 & 0 & 0
	\\
	0 & 0 & 1 & 0
	\\
	0 & 1 & 0 & 0
	\\
	0 & 0 & 0 & 1
	\end{array}
	\right)_{12}
	=
	\left(
	\begin{array}{cccc}
	P^{++}_{++} & P^{++}_{+-} & P^{+-}_{++} & P^{+-}_{+-}
	\\
	P^{++}_{-+} & P^{++}_{--} & P^{+-}_{-+} & P^{+-}_{--}
	\\
	P^{-+}_{++} & P^{-+}_{+-} & P^{--}_{++} & P^{--}_{+-}
	\\
	P^{-+}_{-+} & P^{-+}_{--} & P^{--}_{-+} & P^{--}_{--}
	\end{array}
	\right)_{12}
	=
	P_{12},
	\label{p1}
\end{align}
where $P_{12}$ is the {\it permutation matrix} on the spaces $V_1,V_2$. We shall represent the components of $P_{12}$ by the $P$-vertex
\begin{align}
	(P_{12})^{i_1 j_1}_{i_2 j_2}
	\input{diagrams/define-permutation-tensor}
	\label{p2}
\end{align}
where $i_1,i_2,j_1,j_2$ take values in $\{+,-\}$. Interchanging these indices with their appropriate arrows, the non-zero entries of (\ref{p1}) get paired with $P$-vertices (\ref{p2}) having arrow conservation along each line. In other words,
\begin{align}
	\input{diagrams/permutation-tensor-terms}\ 1
\end{align}
while all remaining $P$-vertices are equal to zero. 

\subsection{Identity matrix $I_{12}$}

The identity matrix $I_{12}$ receives a similar treatment to the preceding matrices. Although its action is trivial, we wish to write it as a vertex in order to place it alongside its companions. Writing
\begin{align}
	I_{12}
	=
	\left(
	\begin{array}{cccc}
	I^{++}_{++} & I^{++}_{+-} & I^{+-}_{++} & I^{+-}_{+-}
	\\
	I^{++}_{-+} & I^{++}_{--} & I^{+-}_{-+} & I^{+-}_{--}
	\\
	I^{-+}_{++} & I^{-+}_{+-} & I^{--}_{++} & I^{--}_{+-}
	\\
	I^{-+}_{-+} & I^{-+}_{--} & I^{--}_{-+} & I^{--}_{--}
	\end{array}
	\right)_{12}
	=
	\left(
	\begin{array}{cccc}
	1 & 0 & 0 & 0
	\\
	0 & 1 & 0 & 0
	\\
	0 & 0 & 1 & 0
	\\
	0 & 0 & 0 & 1
	\end{array}
	\right)_{12}
	\label{i1}
\end{align}
we represent the components of $I_{12}$ by the $I$-vertex
\begin{align}
	(I_{12})^{i_1 j_1}_{i_2 j_2}\ 
	\input{diagrams/define-identity-tensor}
	\label{i2}
\end{align}
where $i_1,i_2,j_1,j_2$ take values in $\{+,-\}$. Like the $P$-vertex, the lines in the $I$-vertex do not genuinely cross. Swapping the indices with their appropriate arrows, the non-zero entries of (\ref{i1}) get paired with $I$-vertices (\ref{i2}) having arrow conservation along each line. That is,
\begin{align}
	\input{diagrams/identity-tensor-trems}\ 1
\end{align}
while all remaining $I$-vertices are equal to zero. At times we shall prefer to represent the identity matrix in the more simple form
\begin{align}
(I_{12})^{i_1 j_1}_{i_2 j_2}\ 
\input{diagrams/define-identity-tensor-alt}
\label{i4}
\end{align}
in which the lines have been explicitly uncrossed. Our preference for a particular notation (\ref{i2}), (\ref{i4}) will depend largely on context.

\subsection{Unitarity relation}\label{2.2}

The $R$-matrix (\ref{r2}) satisfies the unitarity condition
\begin{align}
	R_{21}(\xi_2,\xi_1) R_{12}(\xi_1,\xi_2)
	=
	I_{12} 
	\label{unit1}
\end{align} 
in ${\rm End}(V_1\otimes V_2)$. Alternatively, using diagrammatic tensor notation we can write (\ref{unit1}) as
\begin{align}
	\input{diagrams/unitarity}\ .
	\label{unit2}
\end{align}
Note that here and subsequently, \emph{acting on the left} in the symbolic notation corresponds to \emph{acting on the bottom} in the diagrammatic notation.  Equation (\ref{unit2}) is established by explicitly checking that all 16 components of the tensors agree.

\subsection{Yang-Baxter equation}\label{2.3}

The $R$-matrix (\ref{r2}) is a solution of the Yang-Baxter equation
\begin{align}
	R_{12}(\xi_1,\xi_2) R_{13}(\xi_1,\xi_3) R_{23}(\xi_2,\xi_3)
	=
	R_{23}(\xi_2,\xi_3) R_{13}(\xi_1,\xi_3) R_{12}(\xi_1,\xi_2)
	\label{yb1}
\end{align}
in ${\rm End}(V_1\otimes V_2 \otimes V_3)$. Using diagrammatic tensor notation, equation (\ref{yb1}) takes the form
\begin{align}
	\input{diagrams/yang-baxter}\ .
	\label{yb2}
\end{align}
Equation (\ref{yb2}) is established by explicitly checking that all 64 components of the tensors agree.

\subsection{Monodromy matrices $R_{a,1\ldots n}$ and $R_{1\ldots n,a}$}\label{2.4}

The monodromy matrix $R_{a,1\ldots n}$ is defined as the product
\begin{align}
	R_{a,1\ldots n}(\lambda_a | \xi_1,\ldots,\xi_n)
	&=
	R_{an}(\lambda_a,\xi_n)\ldots R_{a1}(\lambda_a,\xi_1).
	\label{mon1}
\end{align}
The monodromy matrix is a rank $2n+2$ tensor, acting in $V_a\otimes V_1\otimes \cdots \otimes V_n$. Here $V_a$ is another copy of $\mathbb{C}^2$, called the {\it auxiliary space} to distinguish it from the quantum spaces $V_1,\ldots,V_n$. We reserve the variable $\lambda_a$ to accompany the space $V_a$, while the variables $\xi_1,\ldots,\xi_n$ accompany $V_1,\ldots,V_n$ as usual. Often, the $V_a$ dependence of $R_{a,1\ldots n}$ is exhibited explicitly by writing
\begin{align}
	R_{a,1\ldots n}(\lambda_a|\xi_1,\ldots,\xi_n)
	=
	\left(
	\begin{array}{cc}
	A(\lambda_a|\xi_1,\ldots,\xi_n) 
	& 
	B(\lambda_a|\xi_1,\ldots,\xi_n)
	\\
	C(\lambda_a|\xi_1,\ldots,\xi_n) 
	& 
	D(\lambda_a|\xi_1,\ldots,\xi_n)
	\end{array}
	\right)_a
	\label{mon2}
\end{align}
where the entries of the previous matrix are tensors of rank $2n$ acting in $V_1\otimes \cdots \otimes V_n$. The transfer matrix $t(\lambda_a)$, which plays an important role in the algebraic Bethe Ansatz, is defined as
\begin{align}
	t(\lambda_a)
	=
	{\rm tr}_a R_{a,1\ldots n}(\lambda_a|\xi_1,\ldots,\xi_n)
	=
	A(\lambda_a|\xi_1,\ldots,\xi_n)
	+ 
	D(\lambda_a|\xi_1,\ldots,\xi_n).
	\label{t1}
\end{align}
Let us now rephrase these definitions using diagrammatic tensor notation. Firstly, the monodromy matrix (\ref{mon1}) is represented by the string of $R$-vertices
\begin{align}
	R_{a,1\ldots n}(\lambda_a|\xi_1,\ldots,\xi_n)\ 
	\input{diagrams/monodromy-tensor}\ .
	\label{mon3}
\end{align}
A particular component of $R_{a,1\ldots n}(\lambda_a|\xi_1,\ldots,\xi_n)$ is obtained by selecting appropriate values for the indices $i_a,j_a \in \{+,-\}$. For example, to obtain $A(\lambda_a|\xi_1,\ldots,\xi_n)$ one should select $i_a=j_a=+$. Hence we have the diagrammatic representations
\begin{align}
	&\ A(\lambda_a|\xi_1,\ldots,\xi_n) &\ \input{diagrams/monodromy-tensor-A}
	\quad 
	B(\lambda_a|\xi_1,\ldots,\xi_n) &\ \input{diagrams/monodromy-tensor-B}
	\\
	&\ C(\lambda_a|\xi_1,\ldots,\xi_n) &\ \input{diagrams/monodromy-tensor-C}
	\quad
	D(\lambda_a|\xi_1,\ldots,\xi_n) &\ \input{diagrams/monodromy-tensor-D}\ .
	\nonumber
\end{align}
Finally, we notice that the transfer matrix (\ref{t1}) is obtained by identifying the ends of the string of vertices (\ref{mon3}). It follows that
\begin{align}
	t(\lambda_a)
	\ \input{diagrams/transfer}
	\label{t2}
\end{align}
where the marked lines in (\ref{t2}) are taken to be identified.

Before moving on, let us introduce another monodromy matrix whose ordering of spaces is the reverse of (\ref{mon1}). Explicitly speaking, we define 
\begin{align}
	R_{1\ldots n,a}(\xi_1,\ldots,\xi_n|\lambda_a)
	=
	R_{1a}(\xi_1,\lambda_a)\ldots R_{na}(\xi_n,\lambda_a)
	\label{mon4}
\end{align}
and represent it diagrammatically as
\begin{align}
	R_{1\ldots n,a}(\xi_1,\ldots,\xi_n|\lambda_a)
	\ \input{diagrams/monodromy-tensor-r-to-l}\ .
	\label{mon5}
\end{align}
The definitions (\ref{mon4}), (\ref{mon5}) are necessary for the calculations in \S \ref{unit-mon} and \S \ref{cocycle-R}.

\subsection{Unitarity of monodromy matrices}
\label{unit-mon}

Let $R_{a,1\ldots n}(\lambda_a),R_{1\ldots n,a}(\lambda_a)$ denote the monodromy matrices (\ref{mon1}), (\ref{mon4}) respectively. Due to the unitarity relation (\ref{unit1}) they satisfy the equations

\begin{align}
	R_{1\ldots n,a}(\lambda_a)
	R_{a,1\ldots n}(\lambda_a)
	&=
	I_{a,1\ldots n}
	\label{unit3}
	\\
	R_{a,1\ldots n}(\lambda_a)
	R_{1\ldots n,a}(\lambda_a)
	&=
	I_{1\ldots n,a}
	\label{unit5}
\end{align}
where $I_{a,1\ldots n} = I_{an} \ldots I_{a1}$ and $I_{1\ldots n,a} = I_{1a} \ldots I_{na}$ both act identically in $V_a\otimes V_1\otimes \cdots \otimes V_n$. Using diagrammatic notation, (\ref{unit3}) and (\ref{unit5}) may be written as
\begin{align}
	\input{diagrams/monodromy-unitarity-2}
	\label{unit4}
	\\
	\input{diagrams/monodromy-unitarity-1}
	\label{unit6}
\end{align}
which are clearly true by iteration of (\ref{unit2}).

\subsection{Intertwining equation}

Let $R_{a,1\ldots n}(\lambda_a),R_{b,1\ldots n}(\lambda_b)$ be monodromy matrices of the form (\ref{mon1}). By virtue of the Yang-Baxter equation (\ref{yb1}) they satisfy
\begin{align}
	R_{ab}(\lambda_a,\lambda_b)
	R_{a,1\ldots n}(\lambda_a)
	R_{b,1\ldots n}(\lambda_b)
	=
	R_{b,1\ldots n}(\lambda_b)
	R_{a,1\ldots n}(\lambda_a)
	R_{ab}(\lambda_a,\lambda_b).
	\label{int1}
\end{align}
This is known as the intertwining equation. It is important in the algebraic Bethe Ansatz scheme because it generates all commutation relations between the entries of the monodromy matrix (\ref{mon2}). It has the diagrammatic equivalent
\begin{align}
	\input{diagrams/intertwining-1}\ .
	\label{int2}
\end{align}
The proof of equation (\ref{int1}) is simplified if one instead proves its diagrammatic version (\ref{int2}). To be precise, by applying the diagrammatic equation (\ref{yb2}) to the left hand side of (\ref{int2}) $n$ times successively, the right hand side is obtained. The aim of this paper is to extend such efficient diagrammatic proofs to more equations.

\subsection{Alternative version of intertwining equation}
\label{cocycle-R} 

Multiplying (\ref{int1}) by $R_{1\ldots n,b}(\xi_1,\ldots,\xi_n|\lambda_b)$ from the left and right, we obtain
\begin{align}
	R_{1\ldots n,b}(\lambda_b)
	R_{ab}(\lambda_a,\lambda_b)
	R_{a,1\ldots n}(\lambda_a)
	=
	R_{a,1\ldots n}(\lambda_a)
	R_{ab}(\lambda_a,\lambda_b)
	R_{1\ldots n,b}(\lambda_b),
	\label{int3}
\end{align}
which is deduced using the global unitarity relation (\ref{unit3}). In terms of diagrammatic tensors, (\ref{int3}) takes the form
\begin{align}
	\input{diagrams/intertwining-2}\ .
	\label{int4}
\end{align}
We will revisit this identity in \S \ref{cocycle-F}. 

\subsection{$R^{\sigma}_{1\ldots n}$ as a bipartite graph}\label{2.5}

Let $\sigma\{1,\ldots,n\} 
= \{\sigma(1),\ldots,\sigma(n)\}$ be an arbitrary permutation of the set of integers $\{1,\ldots,n\}$. A standard device is to represent this permutation as a bipartite graph. This is achieved by writing down two rows of integers $\{n,\ldots,1\}$ and $\{\sigma(n),\ldots,\sigma(1)\}$, one row directly above the other, and connecting each integer $i$ in the top row with $i$ in the bottom row. The manner in which the lines cross is, for the moment, not important. We denote the resulting graph by $G(\sigma)$. 

Using the diagrammatic representation (\ref{r-vert}) of the $R$-matrix, we define $R^{\sigma}_{1\ldots n}$ to be the rank $2n$ tensor corresponding to the graph $G(\sigma)$. In order to make this identification, the crossing of lines must now be considered. All ways of drawing $G(\sigma)$ are equivalent up to applications of the unitarity (\ref{unit2}) and Yang-Baxter equation (\ref{yb2}). This means that we can prove complicated equations involving products of $R$-matrices by drawing both sides as bipartite graphs and checking that they have the same connectivity from top to base.

To demonstrate this point, consider the example $\sigma\{1,2,3,4,5\} = \{3,5,2,1,4\}$ which is represented by the graphs 
\begin{align}
	G(\sigma)\ \input{diagrams/a-permutation}\ .
	\label{g}
\end{align}
We obtain the tensor $R^{\sigma}_{12345} = R_{25} R_{15} R_{45} R_{12} R_{13} R_{23}$ from the graph on the left, whereas the graph on the right yields $R^{\sigma}_{12345} = R_{25} R_{41} R_{45} R_{15} R_{14} R_{23} R_{13} R_{12}$. These expressions are shown to be equal by application of (\ref{unit1}) and (\ref{yb1}).

When studying the tensors $R^{\sigma}_{1\ldots n}$, whose form can be rather complicated, it is helpful to have a canonical way of breaking the permutation $\sigma$ into elementary steps. To this end, let $\{a_1,\ldots,a_n\}$ be an arbitrary set of distinct integers and define 
\begin{align}
	\sigma_c\{a_1,a_2,\ldots,a_n\}
	=
	\{a_2,\ldots,a_n,a_1\}
	\label{cyc1}
\end{align}
which cyclically permutes the set, and
\begin{align}
	\sigma_{p}\{a_1,a_2,a_3,\ldots,a_n\}
	=
	\{a_2,a_1,a_3,\ldots,a_n\}
	\label{pair1}
\end{align}
which swaps the first pair of the set. Any permutation $\sigma$ of $\{1,\ldots,n\}$ can be written as a composition of the permutations $\sigma_c,\sigma_p$. This means that $R^{\sigma}_{1\ldots n}$ can always be decomposed into a product of tensors of the form $R^{\sigma_c}_{a_1\ldots a_n}$ and $R^{\sigma_p}_{a_1\ldots a_n}$. 

To illustrate this idea, we return to the example $\sigma\{1,2,3,4,5\} = \{3,5,2,1,4\}$. Noticing that this permutation can be expressed as $\sigma_c\circ\sigma_p\circ(\sigma_c)^2\circ\sigma_p\{1,2,3,4,5\}$, we find that
\begin{align}
	R^{\sigma}_{12345}
	=
	R^{\sigma_c}_{43521}
	R^{\sigma_p}_{34521}
	R^{\sigma_c}_{13452}
	R^{\sigma_c}_{21345}
	R^{\sigma_p}_{12345}
\end{align}
or diagrammatically speaking,  
\begin{align}
	R^{\sigma}_{12345}\ \input{diagrams/canonical-decomposition-permutation}\ .
\end{align}
Therefore when proving equations involving $R^{\sigma}_{1\ldots n}$ it is sufficient to verify them for the special cases
\begin{align}
	R^{\sigma_c}_{1\ldots n} \ \input{diagrams/cyclic-permutation} \ R_{1,2\ldots n} 
	\quad\quad
	R^{\sigma_p}_{1\ldots n} \ \input{diagrams/site-swap-permutation} \ R_{1 2}
	\label{cyc-pair}
\end{align}
since $R^{\sigma}_{1\ldots n}$ is generally formed by a large number of such diagrams, up to overall labelling of the lines. We will use this fact frequently throughout the next section.

\section{Graphical construction of $F$-matrices}

\subsection{Elementary factorizing matrix $F_{12}$}\label{3.1}

The simplest factorizing problem consists in finding a matrix $F_{12}(\xi_1,\xi_2)\in {\rm End}(V_1\otimes V_2)$ satisfying
\begin{align}
	F_{21}(\xi_2,\xi_1) R_{12}(\xi_1,\xi_2)
	=
	F_{12}(\xi_1,\xi_2),
	\label{fact1}
\end{align}
where $R_{12}(\xi_1,\xi_2)$ is the $R$-matrix (\ref{r2}). As can be verified by direct calculation of all 16 components in equation (\ref{fact1}), a solution is given by  
\begin{align}
	F_{12}(\xi_1,\xi_2)
	=
	\left(
	\begin{array}{cccc}
	1 & 0 & 0 & 0
	\\
	0 & 1 & 0 & 0
	\\
	0 & c(\xi_1,\xi_2) & b(\xi_1,\xi_2) & 0
	\\
	0 & 0 & 0 & 1
	\end{array}
	\right)_{12}
	\label{f3}
\end{align}
or equivalently,
\begin{align}
	F_{12}(\xi_1,\xi_2)
	&=
	e^{(++)}_1 I_{12} + e^{(--)}_1 R_{12}(\xi_1,\xi_2)
	\label{f4}
	\\
	&=
	e^{(--)}_2 I_{12} + e^{(++)}_2 R_{12}(\xi_1,\xi_2)
	\label{f5}
\end{align}
where we recall that $e^{(++)}_k,e^{(--)}_k \in {\rm End}(V_k)$ denote the matrices
\begin{align}
	e^{(++)}_k
	=
	\left(
	\begin{array}{cc}
	1 & 0
	\\
	0 & 0
	\end{array}
	\right)_k,
	\quad
	e^{(--)}_k
	=
	\left(
	\begin{array}{cc}
	0 & 0
	\\
	0 & 1
	\end{array}
	\right)_k
\end{align}
for $k \in \{1,2\}$. Our first goal is to provide a diagrammatic notation for the matrix (\ref{f3}), which can be extended to more complicated factorizing tensors. For all $i_1,i_2,j_1,j_2 \in \{+,-\}$ we define 
\begin{align}
	(F_{12})^{i_1 j_1}_{i_2 j_2}
	\ \input{diagrams/define-f-tensor-2}\ .
	\label{f-vert1}
\end{align}
The object on the right 	 side of (\ref{f-vert1}) is similar to an $R$-vertex (\ref{r-vert}). The only distinguishing features are the dot $\bullet$ and the triangle $\triangledown$ which have been placed on this vertex. The meaning of these symbols is as follows.
\begin{enumerate}
	\item
		When $i_1 = +$, corresponding to a state arrow that points opposite $\triangledown$, the two lines of the vertex (\ref{f-vert1}) do not genuinely cross. In this case, (\ref{f-vert1}) behaves as an $I$-vertex (\ref{i2}):
\begin{align}
	\input{diagrams/define-f-tensor-3} &\ . 
\end{align}  
	\item
		When $i_1 = -$, corresponding to a state arrow that points with $\triangledown$, the two lines of the vertex (\ref{f-vert1}) cross as they normally would. In this case, (\ref{f-vert1}) behaves as an $R$-vertex (\ref{r-vert}):
\begin{align}
	\input{diagrams/define-f-tensor-4} &\ . 
\end{align}
\end{enumerate}
The diagram (\ref{f-vert1}) is a representation of the equation (\ref{f4}). Observe that it is also possible to define
\begin{align}
	(F_{12})^{i_1 j_1}_{i_2 j_2}
	\ \input{diagrams/define-f-tensor-5}
	\label{f-vert2}
\end{align}
in accordance with equation (\ref{f5}). Either of the representations (\ref{f-vert1}), (\ref{f-vert2}) is valid and we will use them interchangeably, often within the same equation. For example, we propose the equation
\begin{align}
	\input{diagrams/f-tensor-id-1}
	\label{fact2}
\end{align}
in which both representations of the $F$-matrix (\ref{f3}) appear. We can verify that (\ref{fact2}) is true by checking the two cases $i_1 =+$ and $i_1=-$ separately. In the case $i_1 = +$ we have
\begin{align}
	&\input{diagrams/f-tensor-id-1-proof-1}
\end{align}
while in the case $i_1 = -$ we find that
\begin{align}
	&\input{diagrams/f-tensor-id-1-proof-2}\ .
\end{align}
Thus (\ref{fact2}) provides an immediate proof that (\ref{f3}) satisfies (\ref{fact1}). The aim of this section is to simplify the proof of other, more complicated identities by using our adopted notation.    

\subsection{Partial $F$-matrices $F_{1,2\ldots n}$ and $F_{1\ldots n-1,n}$}\label{3.2}

Ultimately we wish to obtain a diagrammatic tensor $F_{1\ldots n}$ which obeys the equation (\ref{fact-big}), where $R^{\sigma}_{1\ldots n}$ is the tensor defined in \S \ref{2.5}. To proceed in this direction, we define the tensors
\begin{align}
	F_{1,2\ldots n}(\xi_1|\xi_2,\ldots,\xi_n) 
	&=
	e^{(++)}_1 
	I_{1,2\ldots n}
	+ 
	e^{(--)}_1 
	R_{1,2\ldots n}(\xi_1|\xi_2,\ldots,\xi_n)
	\label{part1}
	\\
	F_{1\ldots n-1,n}(\xi_1,\ldots,\xi_{n-1}|\xi_n)
	&=
	e^{(--)}_n
	I_{1\ldots n-1,n} 
	+ 
	e^{(++)}_n 
	R_{1\ldots n-1,n}(\xi_1,\ldots,\xi_{n-1}|\xi_n)
	\label{part2}
\end{align}
where $I_{1,2\ldots n} = I_{1n}\ldots I_{12}$ and $I_{1\ldots n-1,n} = I_{1n} \ldots I_{(n-1)n}$ both act identically in $V_1\otimes \cdots \otimes V_n$. In \cite{ms}, these objects were called partial $F$-matrices. When $n=2$ they reduce to the elementary $F$-matrix $F_{12}$.

Let us now extend the diagrammatic notation from \S \ref{3.1} to the tensors (\ref{part1}), (\ref{part2}). We make the identifications
\begin{align}
	F_{1,2\ldots n}(\xi_1|\xi_2,\ldots,\xi_n)
	&\ \input{diagrams/define-mid-f-tensor-1}
	\label{part3}
	\\
	F_{1\ldots n-1,n}(\xi_1,\ldots,\xi_{n-1}|\xi_n)
	&\ \input{diagrams/define-mid-f-tensor-2}\ .
	\label{part4}
\end{align}
The diagrams (\ref{part3}), (\ref{part4}) closely resemble their monodromy matrix counterparts (\ref{mon3}), (\ref{mon5}). The only differences are the arrows $\triangledown,\vartriangle$ and the dots $\bullet$ assigned to every vertex. For the diagram (\ref{part3}), this notation has the following interpretation:
\begin{enumerate}
	\item
		When $i_1=+$, corresponding to a state arrow that points opposite $\triangledown$, the lines in (\ref{part3}) do not genuinely cross. The diagram behaves as a string of $I$-vertices.
\begin{align}
	\input{diagrams/define-mid-f-tensor-3}\ .
\end{align}		
	\item
		When $i_1=-$, corresponding to a state arrow that points with $\triangledown$, the lines in (\ref{part3}) cross normally. The diagram behaves as a string of $R$-vertices; that is, as a monodromy matrix.
\begin{align}
	\input{diagrams/define-mid-f-tensor-4}\ .
\end{align}
\end{enumerate}
An analogous interpretation applies to (\ref{part4}). It is straightforward to check that these conventions are consistent with the algebraic definitions (\ref{part1}) and (\ref{part2}).

\subsection{Cocycle relation}
\label{cocycle-F}

The partial $F$-matrices (\ref{part1}) and (\ref{part2}) satisfy the equation
\begin{align}
	F_{2\ldots n-1,n}(\xi_n)
	F_{1,2\ldots n}(\xi_1)
	=
	F_{1,2\ldots n-1}(\xi_1)
	F_{1\ldots n-1,n}(\xi_n)
	\label{co1}
\end{align}
in ${\rm End}(V_1\otimes\cdots\otimes V_n)$. We prove this identity by writing it diagrammatically as follows,
\begin{align}
	\input{diagrams/cocycle-reln}\ .
	\label{co2}
\end{align}
Here and in subsequent calculations, each row of dots $\bullet$ is associated to its own triangle. For example, studying the left hand side of (\ref{co2}), the top row is sensitive to the state variable in $\triangledown$, while the bottom row is sensitive to the state variable in $\vartriangle$.

In order to prove (\ref{co2}) we isolate its components, and consider each of the cases $\{i_1=+,i_n=+\},\{i_1=+,i_n=-\},\{i_1=-,i_n=+\},\{i_1=-,i_n=-\}$ separately; the remaining spins at $j_1, \dots, j_n$ and $i_2, \dots, i_{n-1}$ remain free. By decomposing (\ref{co2}) in this way we obtain the four equations
\begin{align}
	\input{diagrams/proof-cocycle-reln-2-2}
	\label{co3}
	\\
	\input{diagrams/proof-cocycle-reln-2-1}
	\label{co4}
	\\
	\input{diagrams/proof-cocycle-reln-1-2}
	\label{co5}
	\\
	\input{diagrams/proof-cocycle-reln-1-1}\ .
	\label{co6}   
\end{align}
Equation (\ref{co4}) is true trivially, since in this case none of the lines genuinely cross. Similarly, by unravelling the non-crossing lines in (\ref{co3}) and (\ref{co6}) and using the basic identities
\begin{align}
	\input{diagrams/basic-id-1} \quad\quad\quad \input{diagrams/basic-id-2}
\end{align}
we verify these two cases. The final case (\ref{co5}) is true by using equation (\ref{int4}).

\subsection{Identities involving partial $F$-matrices}\label{3.3}

We now prove a catalogue of identities using our diagrammatic tensor notation. They correspond to equations (56)--(59) in \cite{ms}. In the first two instances the tensor $R^{\sigma}_{1\ldots n}$ is involved, but we shall only give proofs for the special cases $R^{\sigma_c}_{1\ldots n} = R_{1,2\ldots n}$ and $R^{\sigma_p}_{1\ldots n} = R_{12}$. As we explained in \S \ref{2.5}, these specializations are sufficient to prove the general case. 

\myIdentity[1] Let $R^{\sigma}_{1\ldots n}(\xi_1,\ldots,\xi_n)$ be the tensor corresponding to the bipartite graph $G(\sigma)$, and let $F_{a,1\ldots n}(\lambda_a)$ denote a partial $F$-matrix (\ref{part1}) acting in $V_a\otimes V_1\otimes \cdots \otimes V_n$. We have
\begin{align}
	R^{\sigma}_{1\ldots n}(\xi_1,\ldots,\xi_n)
	F_{a,1\ldots n}(\lambda_a)
	=
	F_{a,\sigma(1)\ldots \sigma(n)}(\lambda_a)
	R^{\sigma}_{1\ldots n}(\xi_1,\ldots,\xi_n).
	\label{id1}
\end{align}
\begin{proof} 
	Specializing to the case $\sigma = \sigma_c$, equation (\ref{id1}) becomes
	\begin{align}
		R_{1,2\ldots n}(\xi_1|\xi_2,\ldots,\xi_n)
		F_{a,1\ldots n}(\lambda_a)
		&=
		F_{a,2\ldots n 1}(\lambda_a)
		R_{1,2\ldots n}(\xi_1|\xi_2,\ldots,\xi_n)
	\end{align}
	which may be written in diagrammatic notation as
	\begin{align}
		\input{diagrams/identity-1-a}\ .
		\label{id1-2}
	\end{align}
	We prove the diagrammatic equation (\ref{id1-2}) by considering the cases $i_a=\pm$ separately. When $i_a = +$ all of the vertices marked $\bullet$ become identity matrices, and (\ref{id1-2}) is a trivial statement. When $i_a = -$ only $R$-vertices are present, and (\ref{id1-2}) is true by the intertwining equation (\ref{int2}). 

Specializing to the case $\sigma = \sigma_p$, equation (\ref{id1}) becomes 
\begin{align}
	R_{12}(\xi_1,\xi_2)
	F_{a,1\ldots n}(\lambda_a)
	&=
	F_{a,213\ldots n}(\lambda_a)
	R_{12}(\xi_1,\xi_2)
\end{align}
which may be written in diagrammatic notation as
\begin{align}
	\input{diagrams/identity-1-b}\ .
	\label{id1-3}
\end{align}
When $i_a = +$ all of the vertices marked $\bullet$ become identity matrices, and (\ref{id1-3}) is trivial. When $i_a=-$ only $R$-vertices are present, and (\ref{id1-3}) is true by a single application of the Yang-Baxter equation (\ref{yb2}).
\end{proof}

\myIdentity[2] Let $R^{\sigma}_{1\ldots n}(\xi_1,\ldots,\xi)$ be the tensor corresponding to the bipartite graph $G(\sigma)$, and let $F_{1\ldots n,a}(\lambda_a)$ denote a partial $F$-matrix (\ref{part2}) acting in $V_a\otimes V_1\otimes \cdots \otimes V_n$. We have
\begin{align}
	R^{\sigma}_{1\ldots n}(\xi_1,\ldots,\xi_n)
	F_{1\ldots n,a}(\lambda_a)
	=
	F_{\sigma(1)\ldots \sigma(n),a}(\lambda_a)
	R^{\sigma}_{1\ldots n}(\xi_1,\ldots,\xi_n)\ .
	\label{id2-0}
\end{align}

\begin{proof} The proof is similar to that of identity 1. Specializing to the case $\sigma = \sigma_c$, equation (\ref{id2-0}) becomes 
	\begin{align}
		R_{1,2\ldots n}(\xi_1|\xi_2,\ldots,\xi_n)
		F_{1\ldots n,a}(\lambda_a)
		&=
		F_{2\ldots n 1,a}(\lambda_a)
		R_{1,2\ldots n}(\xi_1|\xi_2,\ldots,\xi_n)
		\label{id2-2}
	\end{align}
	which may be written in diagrammatic notation as
	\begin{align}
		\input{diagrams/identity-2-a}\ .
		\label{id2}
	\end{align}
	We prove the diagrammatic equation (\ref{id2}) by analysing the cases $i_a = \pm$ and using equation (\ref{int4}) in the non-trivial case $i_a=+$. Specializing to the case $\sigma = \sigma_p$, equation (\ref{id2}) becomes
	\begin{align}
		R_{12}(\xi_1,\xi_2)
		F_{1\ldots n,a}(\lambda_a)
		&=
		F_{213\ldots n,a}(\lambda_a)
		R_{12}(\xi_1,\xi_2)
	\end{align}
	which can be written in diagrammatic notation as
	\begin{align}
		\input{diagrams/identity-2-b}
		\label{id2-3}
	\end{align}
and is proved by analysing the cases $i_a = \pm$ and using the Yang-Baxter equation (\ref{yb2}) in the non-trivial case $i_a=+$.
\end{proof}

\myIdentity[3] Let $R_{a,1\ldots n}(\lambda_a)$ denote the monodromy matrix (\ref{mon1}) and similarly let $F_{a,1\ldots n}(\lambda_a), F_{1\ldots n,a}(\lambda_a)$ represent partial $F$-matrices (\ref{part1}), (\ref{part2}). We have
\begin{align}
	F_{1\ldots n,a}(\xi_1,\ldots,\xi_n|\lambda_a)
	R_{a,1\ldots n}(\lambda_a|\xi_1,\ldots, \xi_n)
	=
	F_{a,1\ldots n}(\lambda_a|\xi_1,\ldots,\xi_n)
	\label{id3}
\end{align}

\begin{proof}
	We cast (\ref{id3}) in its diagrammatic form:
	\begin{align}
		\input{diagrams/identity-3}\ .
		\label{id3-2}
	\end{align}
	Once again, we consider the two cases $i_a = \pm$. When $i_a=+$ the vertices on the left hand side of (\ref{id3-2}) are all genuine, while those on the right hand side do not cross. In this case, (\ref{id3-2}) becomes
	\begin{align}
		\input{diagrams/proof-identity-3-1}
		\label{id3-3}
	\end{align}
	which is true by the global unitarity relation (\ref{unit4}). When $i_a=-$ the diagrams for both sides of (\ref{id3-2}) take the same form, and are equal to
	\begin{align}
		\input{diagrams/proof-identity-3-2}\ .
		\label{id3-4}
	\end{align}
	Hence (\ref{id3-2}) is true in both cases.
\end{proof}

\myIdentity[4] Let $R_{1\ldots n,a}(\lambda_a)$ denote the monodromy matrix (\ref{mon4}) and similarly let $F_{a,1\ldots n}(\lambda_a), F_{1\ldots n,a}(\lambda_a)$ represent partial $F$-matrices (\ref{part1}), (\ref{part2}). We have
\begin{align}
	F_{a,1\ldots n}(\lambda_a|\xi_1,\ldots,\xi_n)
	R_{1\ldots n,a}(\xi_1,\ldots, \xi_n |\lambda_a)
	=
	F_{1\ldots n,a}(\xi_1,\ldots, \xi_n |\lambda_a)
	\label{id4}
\end{align}

\begin{proof}
	The proof is very similar to that of identity 3. We write (\ref{id4}) in diagrammatic form:
	\begin{align}
		\input{diagrams/identity-4}\ .
		\label{id4-2}
	\end{align}
	Equation (\ref{id4-2}) is verified by the usual breakdown of the cases $i_a=\pm$.
\end{proof}

\subsection{Full $F$-matrix $F_{1\ldots n}$}
\label{3.4}

We are now in a position to construct solutions to the factorizing equation (\ref{fact-big}). Following \cite{ms}, for all $n \geq 3$ we define the factorizing tensor $F_{1\ldots n}(\xi_1,\ldots,\xi_n) \in {\rm End}(V_1\otimes \cdots \otimes V_n)$ by the recursive equation
\begin{align}
	F_{1\ldots n}(\xi_1,\ldots,\xi_n)
	=
	F_{1\ldots n-1}(\xi_1,\ldots,\xi_{n-1})
	F_{1\ldots n-1,n}(\xi_1,\ldots,\xi_{n-1}|\xi_n)
	\label{f6}
\end{align}
where $F_{1\ldots n-1,n}(\xi_1,\ldots,\xi_{n-1}|\xi_n)$ denotes a partial $F$-matrix of the form (\ref{part2}). The tensors (\ref{f6}) are higher rank extensions of the $F$-matrix $F_{12}(\xi_1,\xi_2) \in {\rm End}(V_1\otimes V_2)$ as defined in \S \ref{3.1}. Solving the recursion (\ref{f6}), we obtain the more explicit formula
\begin{align}
	F_{1\ldots n}(\xi_1,\ldots,\xi_n)
	=
	F_{12}(\xi_1,\xi_2)
	F_{12,3}(\xi_3)
	\ldots
	F_{1\ldots n-1,n}(\xi_n)
	\label{f7}
\end{align}
which is entirely in terms of partial $F$-matrices (\ref{part2}). Using the notation introduced in \S \ref{3.1} and \S \ref{3.2} we are able to obtain a diagrammatic representation of (\ref{f7}). Indeed, equation (\ref{f7}) indicates that we should attach $n-1$ of the diagrams (\ref{part4}) to obtain
\begin{align}
	F_{1\ldots n}(\xi_1,\ldots,\xi_n)
	= \input{diagrams/define-big-f}\ .
	\label{f8}
\end{align}
Notice that we adhere to our previous convention by assuming that all dots $\bullet$ within a row correspond to a single triangle $\vartriangle$.

It is possible to construct an alternative expression for $F_{1\ldots n}(\xi_1,\ldots,\xi_n)$. For all $n \geq 3$ we define the tensor $F'_{1\ldots n}(\xi_1,\ldots,\xi_n) \in {\rm End}(V_1\otimes \cdots \otimes V_n)$ by the recursive equation
\begin{align}
	F'_{1\ldots n}(\xi_1,\ldots,\xi_n)
	=
	F'_{2\ldots n}(\xi_2,\ldots,\xi_n)
	F_{1,2\ldots n}(\xi_1|\xi_2,\ldots,\xi_n)
	\label{f9}
\end{align}
where $F_{1,2\ldots n}(\xi_1|\xi_2,\ldots,\xi_n)$ denotes a partial $F$-matrix (\ref{part1}). Solving the recursion (\ref{f9}), we recover the expression
\begin{align}
	F'_{1\ldots n}(\xi_1,\ldots,\xi_n)
	=
	F_{n-1\ n}(\xi_{n-1},\xi_n)
	\ldots
	F_{2,3\ldots n}(\xi_2)
	F_{1,2\ldots n}(\xi_1)
	\label{f10}
\end{align}
which is entirely in terms of partial $F$-matrices (\ref{part1}). Equivalently, attaching $n-1$ of the diagrams (\ref{part3}) in the manner indicated by (\ref{f10}) we obtain
\begin{align}
	F'_{1\ldots n}(\xi_1,\ldots,\xi_n)
	= \input{diagrams/define-big-f-dash}
	\label{f11}
\end{align}
where all dots $\bullet$ within a row correspond to a single triangle $\triangledown$.

\myIdentity[5] We have
\begin{align}
	F_{1\ldots n}(\xi_1,\ldots,\xi_n)
	&=
	F'_{1\ldots n}(\xi_1,\ldots,\xi_n).
	\label{f12}
\end{align}
\begin{proof}
	The result follows by induction on $n$.  For the base case we recall from \S\ref{3.1} the alternative representations of $F_{12}$,
	\begin{align}
		\input{diagrams/f12-alt-rep}\ .
	\end{align}
	For the inductive step, we start with the diagram (\ref{f8}) for $F_{1\ldots n}$ and again use the alternative representation for $F_{12}$ to obtain
	\begin{align}
		F_{1\ldots n}=\input{diagrams/big-f-equivalence-1}\ .
	\end{align}
	Then by repeated application of the cocycle relation (\ref{co2}), we obtain
	\begin{align}
		F_{1\ldots n}=\input{diagrams/big-f-equivalence-2}
	\end{align}
	thereby completing the required inductive step and establishing (\ref{f12}).
\end{proof}
The equivalence of (\ref{f8}) and (\ref{f11}) will be important in \S \ref{glob-fact}.

\subsection{Global factorizing equation}
\label{glob-fact}

We now prove the result (\ref{fact-big}) for the tensors $R^{\sigma}_{1\ldots n}$ and $F_{1\ldots n}$ as defined in \S \ref{2.5} and \S \ref{3.4}, respectively. Once again we restrict our attention to the cases $R^{\sigma_c}_{1\ldots n} = R_{1,2\ldots n}$ and $R^{\sigma_p}_{1\ldots n} = R_{12}$, which are sufficient.

\myTheorem[1] Let $R^{\sigma}_{1\ldots n}(\xi_1,\ldots,\xi_n)$ be the tensor assigned to the bipartite graph $G(\sigma)$, and $F_{1\ldots n}(\xi_1,\ldots,\xi_n)$ denote the $F$-matrix (\ref{f7}). We have
\begin{align}
	F_{\sigma(1)\ldots\sigma(n)}
	(\xi_{\sigma(1)},\ldots,\xi_{\sigma(n)})
	R^{\sigma}_{1\ldots n}(\xi_1,\ldots,\xi_n)
	=
	F_{1\ldots n}(\xi_1,\ldots,\xi_n).
	\label{fact-big2}
\end{align}

\begin{proof}
In the case $\sigma = \sigma_c$ (recall equation (\ref{cyc1})), equation (\ref{fact-big2}) specializes to
\begin{align}
	F_{2\ldots n 1}(\xi_2,\ldots,\xi_n,\xi_1)
	R_{1,2\ldots n}(\xi_1|\xi_2,\ldots,\xi_n)
	&=
	F_{1\ldots n}(\xi_1,\ldots,\xi_n)
	\label{frf1}
\end{align}
which when written in diagrammatic notation becomes
\begin{align}
	\label{frf1-diag}
	&\ \input{diagrams/big-f-id-1-1} \\ \nopagebreak
	&\ \input{diagrams/big-f-id-1-2}\ . \nonumber
\end{align}
Notice that we have used the representation (\ref{f8}) for $F_{2\ldots n 1}(\xi_2,\ldots,\xi_n,\xi_1)$, whereas the representation (\ref{f11}) has been used for $F_{1\ldots n}(\xi_1,\ldots,\xi_n)$. Treating (\ref{frf1-diag}) as a proposition, we shall proceed from the diagram on its left hand side to the diagram on the right. Firstly, we use equation (\ref{id3-2}) to transform the left hand side into
\begin{align}
	\input{diagrams/proof-big-f-id-1-2}\ .
\end{align}
We then reposition the base of line 1 so that all $I$-vertices are removed from the lattice, in the sense of our earlier remark (\ref{i4}). The result of this elementary transformation is shown below:  
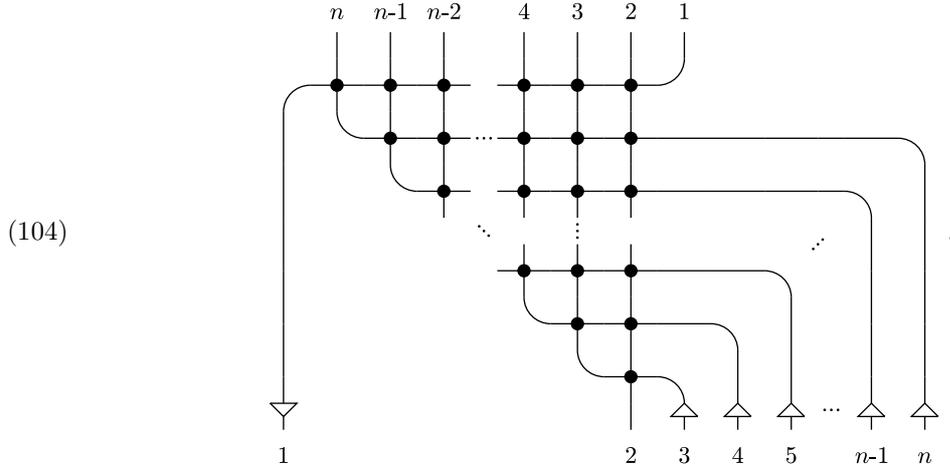
\begin{align}
	\input{diagrams/proof-big-f-id-1-1}\ .
	\label{frf2}
\end{align}
Finally, we use the equivalence between the $F$-matrix representations (\ref{f8}), (\ref{f11}) to transform the remainder of the diagram (\ref{frf2}) into the right hand side of (\ref{frf1-diag}), and the proposition is established.

In the case $\sigma = \sigma_p$, (recall equation (\ref{pair1})) equation (\ref{fact-big2}) specializes to
\begin{align}
	F_{213\ldots n}(\xi_2,\xi_1,\xi_3,\ldots,\xi_n)
	R_{12}(\xi_1,\xi_2)
	&=
	F_{1\ldots n}(\xi_1,\ldots,\xi_n)
	\label{frf4}
\end{align}
which when written in diagrammatic notation becomes
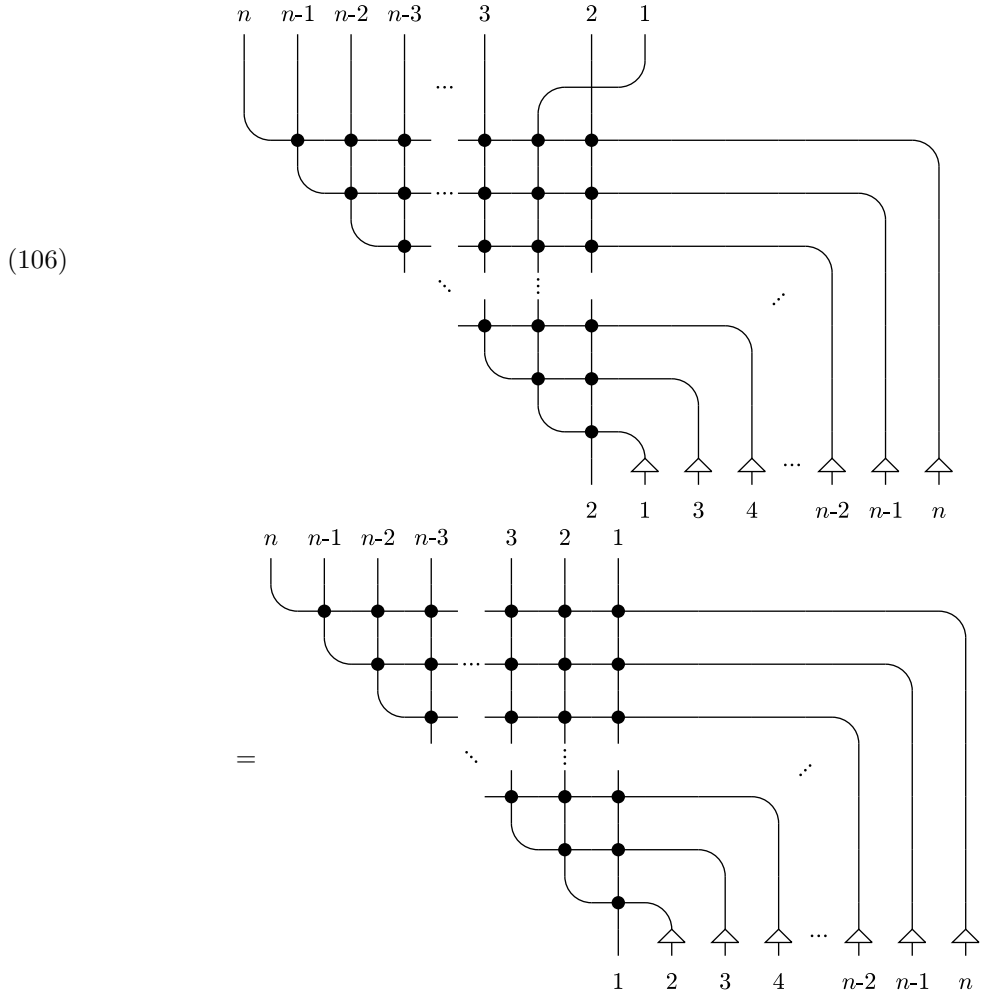
\begin{align}
	\label{frf4-diag}
	&\ \input{diagrams/big-f-id-2-1} \\ \nopagebreak
	&= \input{diagrams/big-f-id-2-2}\ . \nonumber
\end{align}
This time, we use the same representation (\ref{f8}) for both $F_{213\ldots n}(\xi_2,\xi_1,\xi_3,\ldots,\xi_n)$ and $F_{1\ldots n}(\xi_1,\ldots,\xi_n)$. As before, (\ref{frf4-diag}) is considered a proposition and we proceed in steps to show its validity. Firstly, we repeatedly use (\ref{id2-3}) to transform the left hand side of (\ref{frf4-diag}) into
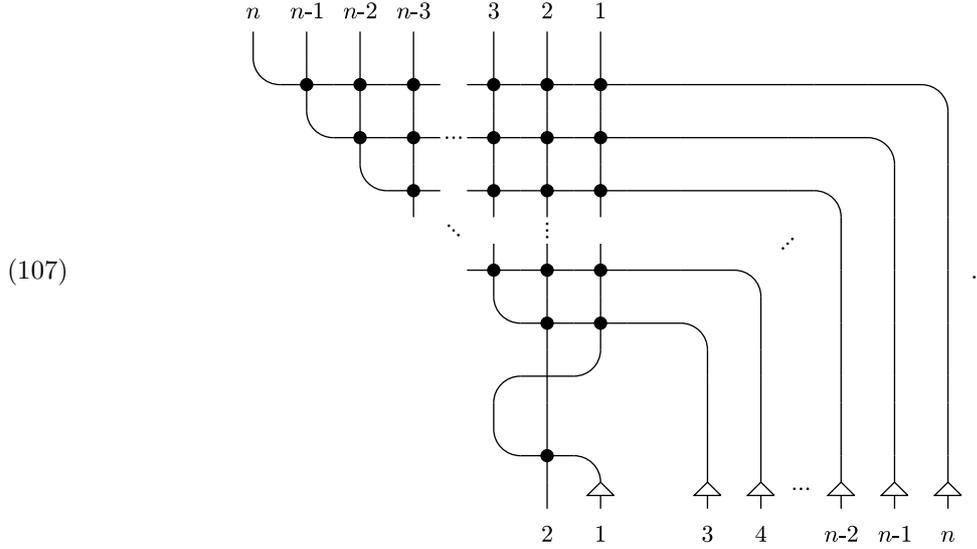
\begin{align}
	\input{diagrams/proof-big-f-id-2-1}\ .
	\label{frf5}
\end{align}
Secondly, applying the relation (\ref{fact2}) to the vertices at the base of (\ref{frf5}) we produce the equivalent diagram
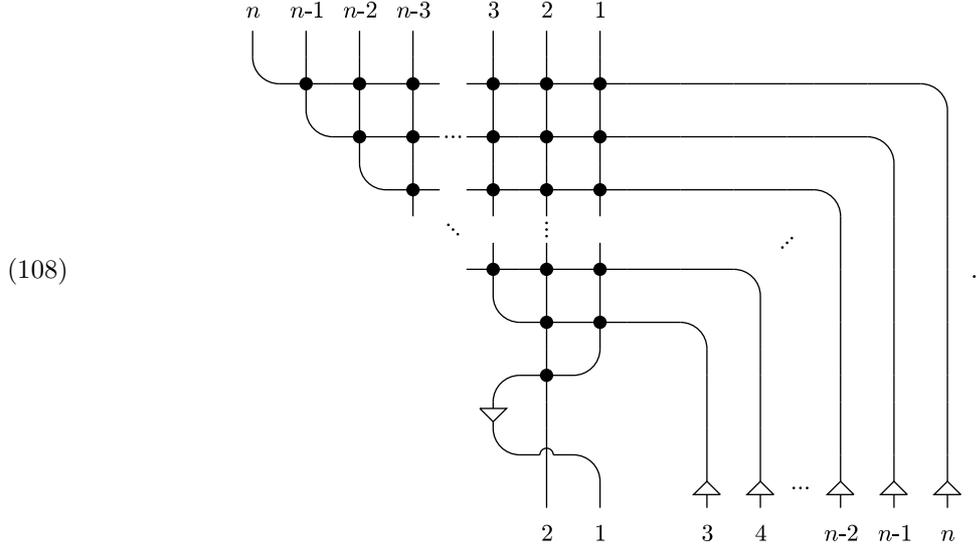
\begin{align}
	\input{diagrams/proof-big-f-id-2-2}\ .
	\label{frf6}
\end{align}
Using (\ref{i4}) to remove the $I$-vertex at the base of (\ref{frf6}), we then apply the equivalence between elementary $F$-matrices (\ref{f-vert1}), (\ref{f-vert2}) to obtain the right hand side of (\ref{frf4-diag}). This proves the proposition. 

The two cases (\ref{frf1}) and (\ref{frf4}) are sufficient to prove (\ref{fact-big2}), since any $R^{\sigma}_{1\ldots n}(\xi_1,\ldots,\xi_n)$ can be decomposed into a product of tensors of the form $R^{\sigma_c}_{a_1\ldots a_n}(\xi_{a_1},\ldots,\xi_{a_n})$ and $R^{\sigma_p}_{a_1 \ldots a_n}(\xi_{a_1},\ldots,\xi_{a_n})$. In conclusion, let us remark that for $2\leq i \leq n$ the relation
\begin{align}
	F_{1\ldots i+1i \ldots n}
	(\xi_1,\ldots,\xi_{i+1},\xi_i,\ldots,\xi_n)
	R_{ii+1}(\xi_i,\xi_{i+1})
	=
	F_{1\ldots n}
	(\xi_1,\ldots,\xi_n)
	\label{frf7}
\end{align}
requires a substantially more involved proof than the two cases above. If we were to prove (\ref{frf7}) directly, we would deconstruct permutations $\sigma$ in terms of pairwise swaps, as is more traditional. However, the simplest proof of (\ref{frf7}) consists of writing $R_{ii+1}$ as a product of $R^{\sigma_c}_{a_1\ldots a_n}$ and $R^{\sigma_p}_{a_1\ldots a_n}$ tensors. Hence for our purposes $\sigma_c$ and $\sigma_p$ comprise the most expedient permutation basis.
\end{proof}

\section{Local spin operators in XXZ spin-$\frac{1}{2}$ chain}

\subsection{Formulae for local spin operators} 
\label{spin}

To begin this section, we list the four formulae which we intend to prove. Recalling the definition of the monodromy matrix operators (\ref{mon2}) and the transfer matrix (\ref{t1}), we claim that 
\begin{align}
	e^{(++)}_i
	&=
	\left(
	\begin{array}{cc}
	1 & 0
	\\
	0 & 0
	\end{array}
	\right)_i
	=
	\left(
	\prod_{j=1}^{i-1}
	t(\xi_j)
	\right)
	A(\xi_i|\xi_1,\ldots,\xi_n)
	\left(
	\prod_{j=i+1}^{n}
	t(\xi_j)
	\right)
	\label{spin1}
	\\
	e^{(-+)}_i
	&=
	\left(
	\begin{array}{cc}
	0 & 0
	\\
	1 & 0
	\end{array}
	\right)_i
	=
	\left(
	\prod_{j=1}^{i-1}
	t(\xi_j)
	\right)
	B(\xi_i|\xi_1,\ldots,\xi_n)
	\left(
	\prod_{j=i+1}^{n}
	t(\xi_j)
	\right)
	\label{spin2}
	\\
	e^{(+-)}_i
	&=
	\left(
	\begin{array}{cc}
	0 & 1
	\\
	0 & 0
	\end{array}
	\right)_i
	=
	\left(
	\prod_{j=1}^{i-1}
	t(\xi_j)
	\right)
	C(\xi_i|\xi_1,\ldots,\xi_n)
	\left(
	\prod_{j=i+1}^{n}
	t(\xi_j)
	\right)
	\label{spin3}
	\\
	e^{(--)}_i
	&=
	\left(
	\begin{array}{cc}
	0 & 0
	\\
	0 & 1
	\end{array}
	\right)_i
	=
	\left(
	\prod_{j=1}^{i-1}
	t(\xi_j)
	\right)
	D(\xi_i|\xi_1,\ldots,\xi_n)
	\left(
	\prod_{j=i+1}^{n}
	t(\xi_j)
	\right).
	\label{spin4}
\end{align}
The local spin operators are given by the equations $S_i^{+} = e^{(+-)}_i, S_i^{-} = e^{(-+)}_i, S_i^{z} = \frac{1}{2}(e^{(++)}_i-e^{(--)}_i)$. These formulae were originally derived in \cite{kmt}. Since they embed local spin operators in the algebra of monodromy matrix entries, they have proved useful in the calculation of correlation functions \cite{kmt2,kmst}.

\subsection{An identity involving $R^{\sigma}_{1\ldots n}$}\label{4.1}

We present a result which parallels identity 1 from \S \ref{3.3}. This result is important when studying the properties of the monodromy matrix in the $F$-basis.

\myIdentity[6] Let $R^{\sigma}_{1\ldots n}(\xi_1,\ldots,\xi_n)$ be the tensor corresponding to the bipartite graph $G(\sigma)$, and let $R_{a,1\ldots n}(\lambda_a)$ denote the monodromy matrix (\ref{mon1}). We have
\begin{align}
	R^{\sigma}_{1\ldots n}(\xi_1,\ldots,\xi_n)
	R_{a,1\ldots n}(\lambda_a)
	=
	R_{a,\sigma(1)\ldots\sigma(n)}(\lambda_a)
	R^{\sigma}_{1\ldots n}(\xi_1,\ldots,\xi_n).  
	\label{rr1}
\end{align}

\begin{proof} 
	The proof is accomplished by verifying (\ref{rr1}) for the particular permutations $\sigma=\sigma_c,\sigma_p$. In fact we have already done this in the proof of identity 1, when considering the $(i_a=-)$ case. For $\sigma = \sigma_c$ we have  
	\begin{align}
		R_{1,2\ldots n}(\xi_1|\xi_2,\ldots,\xi_n)
		R_{a,1\ldots n}(\lambda_a)
		&=
		R_{a,2\ldots n 1}(\lambda_a)
		R_{1,2\ldots n}(\xi_1|\xi_2,\ldots,\xi_n)
		\label{rr2} 
	\end{align}
	which in diagrammatic notation may be written
	\begin{align}
		\input{diagrams/intertwining-2-alt} \nonumber
	\end{align}
	while for the $\sigma = \sigma_p$ case we have
	\begin{align}
		R_{12}(\xi_1,\xi_2)
		R_{a,1\ldots n}(\lambda_a)
		&=
		R_{a,2 1 3\ldots n}(\lambda_a)
		R_{12}(\xi_1,\xi_2) 
		\label{rr3}
	\end{align}
	which in diagrammatic notation may be written
	\begin{align}
		\input{diagrams/rr-identity}\ . \nonumber
	\end{align}
	The equations (\ref{rr2}) and (\ref{rr3}) are sufficient to prove (\ref{rr1}). 
\end{proof}

The significance of (\ref{rr1}) is demonstrated by using the factorization (\ref{fact-big2}) of $R^{\sigma}_{1\ldots n}$. Making this substitution, we find that
\begin{align}
	F_{1\ldots n} 
	R_{a,1\ldots n}(\lambda_a) 
	F_{1\ldots n}^{-1}
	=
	F_{\sigma(1)\ldots \sigma(n)}
	R_{a,\sigma(1)\ldots \sigma(n)}(\lambda_a)
	F_{\sigma(1)\ldots \sigma(n)}^{-1}.
	\label{twist}
\end{align} 
Defining the {\it twisted} monodromy matrix $\widetilde{R}_{a,1\ldots n}$ by
\begin{align}
	\widetilde{R}_{a,1\ldots n}(\lambda_a)
	=
	F_{1\ldots n}(\xi_1,\ldots,\xi_n)
	R_{a,1\ldots n}(\lambda_a)
	F_{1\ldots n}^{-1}(\xi_1,\ldots,\xi_n)
\end{align}
we see from (\ref{twist}) that it satisfies
\begin{align}
	\widetilde{R}_{a,1\ldots n}(\lambda_a)
	=
	\widetilde{R}_{a,\sigma(1)\ldots\sigma(n)}(\lambda_a).
\end{align}
Hence the twisted monodromy matrix $\widetilde{R}_{a,1\ldots n}$ is invariant under permutations of its quantum spaces. This observation was used in \cite{kmt} to construct the formulae (\ref{spin1})--(\ref{spin4}). Our derivation of these formulae will be based solely on (\ref{rr2}).

\subsection{Spin chain propagator $U_p^{q}$}\label{4.2}

Examining equation (\ref{rr2}), we see that $R_{1,2\ldots n}$ cyclically permutes the quantum spaces of the monodromy matrix $R_{a,1\ldots n}$. Thus we say that it propagates the spin chain from site 1 to site 2. By extension of this argument, for all $1\leq p < q \leq n$ the tensor  
\begin{align}
	U_p^{q}
	=
	R_{q-1,q\ldots n 1\ldots q-2}(\xi_{q-1})
	\ldots
	R_{p,p+1\ldots n 1 \ldots p-1}(\xi_p)
	\label{prop1}
\end{align}
propagates the spin chain from site $p$ to site $q$. We call $U_{p}^{q}$ the {\it spin chain propagator.} In diagrammatic tensor notation it is given by
\begin{align}
	U_p^{q}&\ \input{diagrams/defineupq}.
	\label{prop2}
\end{align}
Alternatively, $U_{p}^{q}$ admits an equivalent diagrammatic representation as a rectangular lattice:
\begin{align}
	U_p^{q}&\ \input{diagrams/defineupq-simplified}\ .
	\label{prop7}
\end{align}
To prove that the diagrams (\ref{prop2}) and (\ref{prop7}) are equal, we resort to the idea explained in \S\ref{2.5}, and consider both as bipartite graphs. We immediately see that each diagram has the same connectivity from top to base, proving that they are equal up to applications of the unitarity (\ref{unit2}) and Yang-Baxter equation (\ref{yb2}).  

Now let $U_1^1$ denote the propagator through the entire chain, from site 1 to site $1\ {\rm mod}\ n$. We claim that 
\begin{align}
	U_1^1
	=
	R_{n,1\ldots n-1}(\xi_n)
	\ldots
	R_{1,2\ldots n}(\xi_1)
	\label{prop3}
\end{align}
acts identically in $V_1 \otimes \cdots \otimes V_n$, or equivalently,  
\begin{align}
	U_1^1 &\ \input{diagrams/defineu11-alt}\ .
\end{align}
Once again, this statement is obvious when the diagrams are considered as bipartite graphs. In either diagram, for all $1\leq i \leq n$, line $i$ attaches to the top and base in the position which is $i$ steps from the right. Hence the connectivity of these graphs is the same, and they are necessarily equal. 

\subsection{Monodromy matrix under spin chain propagation}\label{4.3}

We prove, diagrammatically, the equation for the spin-chain propagation of the monodromy matrix. 

\myIdentity[7] Let $R_{a,1\ldots n}$ denote the monodromy matrix (\ref{mon1}), and for all $1\leq i \leq n$ let
\begin{align}
	U_1^{i}
	&=
	R_{i-1,i\ldots n 1\ldots i-2}(\xi_{i-1})
	\ldots
	R_{1,2\ldots n}(\xi_1)
	\\
	U_{i}^1
	&=
	R_{n,1\ldots n-1}(\xi_n)
	\ldots
	R_{i,i+1\ldots n 1\ldots i-1}(\xi_{i})
\end{align}
be propagators from site 1 to site $i$, and from site $i$ to site $1\ {\rm mod}\ n$, respectively. They satisfy the equation
\begin{align}
	U_1^{i}
	R_{a,1\ldots n}(\lambda_a|\xi_1,\ldots,\xi_n)
	U_{i}^1
	=
	R_{a,i\ldots n 1\ldots i-1}
	(\lambda_a|\xi_{i},\ldots,\xi_n,\xi_1,\ldots,\xi_{i-1}).
	\label{id6}
\end{align}
\begin{proof}
	In diagrammatic notation, the proposed equation (\ref{id6}) takes the form
	\begin{align}
		\input{diagrams/rrrrarrr-1}
		\label{id6-2}
	\end{align}
	\begin{align}
		\input{diagrams/rrrrarrr-2}\ . \nonumber
	\end{align}
Comparing the diagrams on either side of (\ref{id6-2}), it is clear that they have the same connectivity from top to base. This shows that they are equal, due to the reasons already discussed.
\end{proof} 

\subsection{$U_p^{q}$ as product of transfer matrices}\label{4.4}

The next step of our calculations is the conversion of $U_p^q$ into a product of transfer matrices (\ref{t1}). To do this, we require the following result.

\myIdentity[8] For all $1\leq i\leq n$ we claim that
\begin{align}
	t(\xi_i)
	=
	R_{i,i+1\ldots n 1\ldots i-1}
	(\xi_i|\xi_{i+1},\ldots,\xi_n, \xi_1,\ldots,\xi_{i-1})
\end{align}
where $t(\xi_i)$ is the transfer matrix (\ref{t1}) evaluated at $\lambda_a = \xi_i$.
\begin{proof}
	We start from the diagram (\ref{t2}) for the transfer matrix, and set $\lambda_a = \xi_i$. This produces a $P$-vertex at the intersection of lines $a,i$:  	
\begin{align}
		t(\lambda_a)\Big|_{\lambda_a = \xi_i}
		\ \input{diagrams/transfer-2}\ .
		\label{mon6}
\end{align}
Recall from \S \ref{perm} that all non-zero $P$-vertices exhibit arrow conservation along their lines. It follows that the $P$-vertex may be deleted from the diagram (\ref{mon6}), leaving it invariant:
\begin{align}
		t(\lambda_a)\Big|_{\lambda_a = \xi_i}
		\ \input{diagrams/transfer-3}\ .
		\label{mon7}
\end{align}
Finally, by performing a trivial rotation of the right hand side of (\ref{mon7}) we obtain the equivalent relation
	\begin{align}
		t(\lambda_a)\Big|_{\lambda_a = \xi_i}
		\input{diagrams/transfer-4}
		= 
		R_{i,i+1\ldots n 1\ldots i-1}
		(\xi_i|\xi_{i+1},\ldots,\xi_n, \xi_1,\ldots,\xi_{i-1}).
	\end{align} 
\end{proof}

By virtue of identity 8 and the definition (\ref{prop1}), we are now able to write
\begin{align}
	\prod_{j=p}^{q-1}
	t(\xi_j)
	=
	U_p^q	
	\label{prop5}
\end{align}
where the ordering of the product (\ref{prop5}) is rendered irrelevant, since the transfer matrices $t(\lambda_a),t(\lambda_b)$ commute for all $\lambda_a,\lambda_b$. 

\subsection{Construction of local spin operators}\label{4.5}

We are now in a position to prove the formulae listed in \S \ref{spin}, using diagrammatic tensor notation. We begin by proving identity 9 and then show that it specializes to each of (\ref{spin1})--(\ref{spin4}).

\myIdentity[9] For all $1\leq j \leq n$, let $t(\xi_j)$ be the transfer matrix (\ref{t1}) evaluated at $\lambda_a = \xi_j$. Similarly let $R_{a,1\ldots n}(\xi_i)$ denote the monodromy matrix (\ref{mon1}) evaluated at $\lambda_a = \xi_i$. We have the identity
\begin{align}
	\left(
	\prod_{j=1}^{i-1}
	t(\xi_j)
	\right)
	R_{a,1\ldots n}(\xi_i)
	\left(
	\prod_{j=i+1}^{n}
	t(\xi_j)
	\right)
	=
	P_{ai}
	\label{id8}
\end{align}
which holds in ${\rm End}(V_a\otimes V_1 \otimes \cdots \otimes V_n)$.

\begin{proof} 
Using the result of the previous subsection, we know that 
\begin{align}
	\prod_{j=1}^{i-1} t(\xi_j)
	=
	U^{i}_{1},
	\quad
	\prod_{j=i+1}^{n} t(\xi_j)
	=
	U^{1}_{i+1}.
\end{align}
This means that the left hand side of (\ref{id8}) can be cast into the following diagrammatic form:
\begin{align}
	&
	\left(
	\prod_{j=1}^{i-1}
	t(\xi_j)
	\right)
	R_{a,1\ldots n}(\lambda_a)
	\left(
	\prod_{j=i+1}^{n}
	t(\xi_j)
	\right)
	=
	\label{id8-2}
	\\
	&
	\input{diagrams/p-1}\ .
	\nonumber
\end{align}
Treating (\ref{id8-2}) as a bipartite graph, we notice that it is equivalent to the much simpler diagram
\begin{align}
	\left(
	\prod_{j=1}^{i-1}
	t(\xi_j)
	\right)
	R_{a,1\ldots n}(\lambda_a)
	\left(
	\prod_{j=i+1}^{n}
	t(\xi_j)
	\right)
	= 
	\input{diagrams/p-4}
\end{align}
which has the same connectivity from top to base. Now by setting the variable $\lambda_a$ equal to $\xi_i$, the intersection of the $a,i$ lines splits into a $P$-vertex:
\begin{align}
	\left(
	\prod_{j=1}^{i-1}
	t(\xi_j)
	\right)
	R_{a,1\ldots n}(\xi_i)
	\left(
	\prod_{j=i+1}^{n}
	t(\xi_j)
	\right)
	= 
	\input{diagrams/p-5}.
	\label{id8-3} 
\end{align}
Finally, applying the global unitarity relation (\ref{unit6}) to the left-most portion of (\ref{id8-3}) we are able to write
\begin{align}
	\left(
	\prod_{j=1}^{i-1}
	t(\xi_j)
	\right)
	R_{a,1\ldots n}(\xi_i)
	\left(
	\prod_{j=i+1}^{n}
	t(\xi_j)
	\right)
	&= 
	\input{diagrams/p-6} \label{id8-4}
   \\
	&=
	\input{diagrams/p-7}\ \nonumber
	\\
	&=
	P_{ai}
	\nonumber
\end{align}
which completes the proof of (\ref{id8}).\footnote{Actually, the leap from (\ref{id8-3}) to (\ref{id8-4}) does not constitute a {\it direct} application of (\ref{unit6}), since there is a $P$-vertex present which permutes the positions of the $a,i$ labels at the top of the diagram. However, when one considers that the right-most line is effectively disjoint from the diagram, while the labels are only superficial bookmarks, this step is easily deduced from (\ref{unit6}).}

\end{proof}
Extracting the ${\rm End}(V_a)$ dependence from (\ref{id8}), we recover each of the equations (\ref{spin1})--(\ref{spin4}).

\section{Discussion}

Following the algebraic methods of \cite{ms}, we have outlined a diagrammatic treatment of the factorizing $F$-matrices. The main feature of our work is the diagrammatic depiction of the partial $F$-matrices $F_{a,1\ldots n}$ and $F_{1\ldots n, a}$ in \S\ref{3.2}, which parallel the standard representation of the XXZ monodromy matrix. In \S\ref{3.3} we gave diagrammatic proofs of a number of identities involving partial $F$-matrices. These proofs are quite transparent in our notation, which allows the components of all tensors to be extracted automatically. In \S\ref{3.4} we built the full $F$-matrix $F_{1\ldots n}$ by stacking partial $F$-matrices together and proved the factorizing equation (\ref{fact-big}) in the sufficient cases $\sigma =\sigma_{c}$ and $\sigma =\sigma_p$. Our proofs are inductive in nature, since they only require iterations of the more basic identities derived in \S\ref{3.3}.

We also considered the problem of constructing the local spin operators $S_i^{\pm}$ and $S_i^{z}$ in terms of monodromy matrix elements, which was solved in \cite{kmt}. Using the diagrammatic representation of $U_1^{i}$, in \S\ref{4.3} we derived an equation for the monodromy matrix under spin chain propagation. In \S\ref{4.5} the reconstruction formulae for the local spin operators were derived diagrammatically. Factorizing matrices do not appear explicitly in any of our calculations in \S\ref{4.5}, so our construction of the local spin operators is slightly more direct than that of \cite{kmt}.

\section*{Acknowledgments}

We would like to thank Omar Foda for many valuable discussions and for reading the manuscript. SM is supported by a Melbourne Research Scholarship. MW is supported by a Research Assistantship in the Department of Mathematics and Statistics, University of Melbourne.  This work was produced using free, open-source software, in particular the diagrams were created in \emph{Inkscape}; we would like to acknowledge the thousands of people who have contributed to the development of this software.

\end{document}